\documentclass[11pt]{article}
\usepackage[a4paper,margin=24mm]{geometry}
\usepackage{amsmath,amssymb,amsthm,mathtools,microtype,booktabs}
\usepackage{graphicx}
\usepackage[hidelinks]{hyperref}
\hypersetup{pdftitle={Finite-time negativity freezing dynamics in 4 x 4 quantum systems},pdfauthor={Wenlin Zhang; Mazhar Ali; Haobo Ma},pdfkeywords={Negativity; Entanglement freezing; Collective dephasing; Partial transpose; Four-level systems}}
\newtheorem{theorem}{Theorem}
\newtheorem{lemma}{Lemma}
\newcommand{\ket}[1]{\lvert#1\rangle}
\newcommand{\bra}[1]{\langle#1\rvert}
\newcommand{\pr}[1]{\ket{#1}\bra{#1}}
\newcommand{\diag}{\operatorname{diag}}
\newcommand{\rank}{\operatorname{rank}}
\newcommand{\Tr}{\operatorname{Tr}}
\newtheorem{proposition}{Proposition}
\newcommand{\PT}{\mathsf T_B}
\title{Finite-time negativity freezing dynamics in $4 \otimes 4$ quantum systems}
\author{Wenlin Zhang$^{1,2}$, \qquad Mazhar Ali$^3$ \qquad Haobo Ma$^2$} 
\date{\small
\begin{minipage}{0.96\textwidth}\centering
$^1$National University of Singapore, 117417 Singapore\\[5pt]
$^2$ChronoAI Pte Ltd, Singapore\\
$^3$Department of Electrical Engineering, Faculty of Engineering,\\
Islamic University of Madinah, 107 Madinah, Saudi Arabia\\[3pt]
\href{mailto:e1327962@u.nus.edu}{e1327962@u.nus.edu}\\[3pt]
\href{mailto:mazhar.ali@iu.edu.sa}{mazhar.ali@iu.edu.sa}\quad
\href{mailto:auric@aelf.io}{auric@aelf.io}\\
W. Zhang: \href{https://orcid.org/0009-0005-2414-7968}{ORCID}\qquad
M. Ali: \href{https://orcid.org/0000-0002-3708-1970}{ORCID}\qquad
H. Ma: \href{https://orcid.org/0009-0008-4944-977X}{ORCID}\\[3pt]
Corresponding author: Wenlin Zhang, \href{mailto:e1327962@u.nus.edu}{e1327962@u.nus.edu}
\end{minipage}\\[10pt]22 November 2026}
\begin{document}
\maketitle

\begin{abstract}
We classify exact finite-time negativity freezing for a two-parameter family on $\mathbb C^4\otimes\mathbb C^4$ under shared local Gaussian dephasing. On the equal-amplitude slice, $w=1$, a strictly positive plateau is attained after a finite transient if and only if $1/4<m<1$, and its value is $(4m-1)/12$. For $0\leq m<1/4$, negativity is strictly decreasing at every finite time and reaches its positive limit only asymptotically; it vanishes identically at $m=1/4$, while the endpoint $m=1$ also has no finite-time plateau. The proof isolates an invariant partial-transpose eigenvalue and five complementary three-dimensional blocks. An equality-at-two-times lemma excludes temporary flat intervals, and block positivity gives an exact freezing-boundary criterion. Writing $q=e^{-\kappa t}$, with $\kappa>0$, the excess above the limit for fixed $0<m<1/4$ is of order $q^{32}$, explaining why strict asymptotic decay can appear frozen in ordinary plots; at $m=0$ the leading order is $q^2$. An explicit separability certificate completes the initial-state classification. Supporting calculations give a Gaussian Kraus representation, comparisons with a distinct global channel and known states, and an exact application of an existing realignment-moment criterion. The channel distinction follows our earlier cyclic two-qutrit analysis \cite{zam}; the result here is the finite-time versus asymptotic classification for the specified four-level family and additive local generator.
\end{abstract}


\section{Introduction}
We determine when shared local Gaussian dephasing makes the negativity of an evolving two-four-level state exactly constant after a finite time. For the family introduced below, the equal-amplitude slice has a positive finite-time plateau exactly when $1/4<m<1$; the lower-weight states approach their limiting negativity strictly asymptotically. Throughout this paper, entanglement freezing means freezing of negativity, not constancy of every entanglement measure. We distinguish it from time-invariant negativity, which remains constant from the initial time while the state changes.

High-dimensional entanglement provides resources for quantum information \cite{Erhard-NRP-2020}, but its detection and its evolution under noise depend on both the state and the channel \cite{Friis-2019,Horodecki-RMP81,Guehne-PR474}. Decoherence can cause asymptotic decay or finite-time disentanglement \cite{Yu-work-2,Yu-work-3}; collective noise also permits entanglement to persist while the state evolves \cite{Carnio-PRL115,Carnio-NJP18}. For negativity, an apparently flat late-time curve does not distinguish exact finite-time freezing from a small asymptotic correction. That distinction requires control of the partial-transpose spectrum.

The positive-partial-transpose (PPT) condition is necessary for separability \cite{peres,horodecki} and sufficient in $2\otimes2$ and $2\otimes3$, but not in general in $4\otimes4$. We prove that our family has exactly one PPT member and give its product-state decomposition. Negativity \cite{vidal} quantifies the negative partial-transpose spectrum and vanishes on PPT states. Comparisons with realignment-based tests \cite{rudolph,chen,aam} and a known rank-six PPT-entangled state \cite{tv} provide supporting checks on the initial family.

The charge-sector description of collective dephasing is established in the work of Carnio, Buchleitner, and Gessner \cite[Eqs.~(2)--(5)]{Carnio-PRL115}; phase-damping Kraus representations are also discussed by Liu et al.\ \cite[Eqs.~(44)--(51)]{liu}. The generator matters: shared local random-unitary noise preserves separability, whereas a diagonal global generator need not do so. Our earlier cyclic two-qutrit preprint \cite{zam} already distinguishes the global and correlated-local channels and demonstrates entanglement creation by the global map. It also gives exact PPT/separability regions, product certificates and time-invariant-negativity regimes. On its selected family, the three-sector global map agrees with a correlated local random-unitary semigroup, although the channels differ on arbitrary inputs. The present family has different support and an equally spaced additive local generator with seven charge sectors. Its positive plateau is reached after a transient, and the proof also identifies the parameters for which the limit is never attained at finite time.

Ali's qutrit--qutrit study \cite{aliqutrit} reports freezing through a fixed negative partial-transpose eigenvalue with changing complementary contributions. Here we control the signs of all complementary blocks under the specified shared-local channel, yielding an exact attainment criterion. A qubit--qutrit example in \cite[Section~IV.2, Eq.~(28), p.~7 of the arXiv v1 PDF]{ali2019} instead has the eigenvalue
\[
v_2=\frac{\alpha-\sqrt{\alpha^2+4\xi^{18}(1-\alpha)^2}}{4}.
\]
For $0<\alpha<1$, it is negative at every finite-time value $\xi>0$ and tends to zero only as $\xi\to0$. For that particular example, convergence is therefore asymptotic in the terminology adopted here. This comparison concerns the displayed example; the state families and generators in the cited studies differ from those below.

Tchoffo et al.\ \cite{tchoffo2019} study one-parameter qubit--qutrit states under independent and common classical static, Ornstein--Uhlenbeck and random telegraph noise. They report frozen negativity, described as perfect protection from the common noise, for suitable initial parameters, while independent noise gives finite-time loss of entanglement. Their noise models and protected regimes differ from the shared Brownian-phase Gaussian semigroup considered here. Our result classifies exact equality to the limiting negativity after a finite transient and proves strict asymptotic convergence in the complementary lower-weight regime, using the partial-transpose blocks and an equality-at-two-times argument.

Theorem~\ref{freezing} is the main result. Its proof reduces the $16\times16$ partial transpose to an invariant eigenvalue and five complementary three-dimensional blocks, treats all weight endpoints, and gives an exact boundary criterion through block positivity. The equality-at-two-times lemma turns the strict inequality above the limit into strict monotonicity before freezing, or at every finite time in the asymptotic regime. The $q^{32}$ tail for fixed $0<m<1/4$ makes the distinction relevant even when numerical trajectories appear constant.

The remaining calculations support this classification. We compare the shared-local channel with the distinct three-sector global map, evaluate the existing partial-realignment-moment criterion of Aggarwal, Adhikari, and Majumdar, and identify a local phase-averaging connection with known circulant states \cite{ck,jba}. The explicit seven-operator Kraus list specializes the Rogers--Szeg\H{o} coefficients of Jang and Song \cite[Proposition~4.13]{js}; the Appendix supplies a self-contained factorization and endpoint normalization. Section~\ref{sc} introduces the states and channels, Section~\ref{dynamics} proves the freezing classification, and the subsequent sections contain the known-state and criterion comparisons.

\section{The states and channels}\label{sc}
We introduce a two-parameter family of quantum states described by
\begin{equation}
\rho_{m,w} = m \, \ket{\Phi}\bra{\Phi} +  \frac{1-m}{4} \, \sum_{j=0}^{3}\ket{\chi_j(w)}\bra{\chi_j(w)},
 \label{state}
\end{equation}
where $0\le m\le1,\, w \ge 0$ and the five vectors are given as 
\begin{align}
\ket\Phi&=\tfrac12(\ket{01}+\ket{12}+\ket{23}+\ket{30}),\notag\\
\ket{\chi_0}&=(w \, \ket{00}+\ket{13}+\ket{21})/\sqrt{w^2+2},\notag\\
\ket{\chi_1}&=(w \, \ket{11}+\ket{20}+\ket{32})/\sqrt{w^2+2},\notag\\
\ket{\chi_2}&=(w \, \ket{22}+\ket{03}+\ket{31})/\sqrt{w^2+2},\notag\\
\ket{\chi_3}&=(w \, \ket{33}+\ket{02}+\ket{10})/\sqrt{w^2+2}.
\label{vectors}
\end{align}
The five vectors have disjoint supports and are orthonormal. In particular Eq.(\ref{state}) is a density matrix, of rank five for $0<m<1$.
The basis labels listed in \eqref{vectors}, including the zero-amplitude labels when $w=0$, partition the product basis into modular phase classes. At the balanced point these classes are the actual supports and yield an explicit product decomposition. The same sparse pattern reduces partial transposition to small invariant blocks, which organize the dynamical analysis below.
We use $\mathcal N(\rho)=(\|\rho^{\PT}\|_1-1)/2$ throughout. This family has negative partial transpose (NPT) everywhere except at one specific point, that is, the partial transpose is positive if and only if $(m,w) = (1/4,1)$. We prove separability at this PPT point below. 

\begin{proposition}\label{ppt-classification}
The state $\rho_{m,w}$ is PPT exactly at $(m,w)=(1/4,1)$, where it is separable. Every other member of \eqref{state} has a negative
partial transpose.
\end{proposition}
\begin{proof}
At $m=1$ the state is the maximally entangled pure state $\Phi$, with partial-transpose eigenvalue $-1/4$.
For $m<1$, put $a=m/4$ and $b=(1-m)/(4(w^2+2))>0$. The principal block of $\rho_{m,w}^{T_B}$ on $03,10,21,32$ is
\[
\begin{pmatrix}
b&b\, w &a&b\, w\\
b\, w&b&b\, w&a\\
a&b\, w&b&b\, w\\
b\, w&a&b\, w&b
\end{pmatrix}.
\]
Its eigenvalues include $b-a$ twice and $a+b-2bw$, as is seen on the vectors $(1,0,-1,0)$, $(0,1,0,-1)$ and $(1,-1,1,-1)$.
The principal block on $00,12$ is $\left(\begin{smallmatrix}b\, w^2&b\\b&a\end{smallmatrix}\right)$.
Writing $r=a/b$, positivity of these blocks requires
\[
r\leq1,\qquad 2\, w\leq r+1\leq2,
\qquad r\, w^2\geq1.
\]
Since $r\geq0$ and $w\geq0$, these inequalities force $r=w=1$, and hence $m=1/4$.

For the converse, let $\xi=e^{2\pi i/5}$ and $(r_0,r_1,r_2,r_3)=(0,3,4,1)$, $(c_0,c_1,c_2,c_3)=(0,1,3,2)$.
The normalized vectors
\[
\ket{x_k}=\frac12\sum_{i=0}^3\xi^{kr_i}\ket i,\qquad
\ket{y_k}=\frac12\sum_{j=0}^3\xi^{kc_j}\ket j
\quad(0\leq k\leq4)
\]
give the explicit decomposition
\begin{equation}
\rho_{1/4,1}=\frac15\sum_{k=0}^4
\pr{x_k}\otimes\pr{y_k}.\label{five-product}
\end{equation}
Indeed the average retains the matrix unit $\ket{ij}\bra{p\ell}$ exactly when $r_i+c_j=r_p+c_\ell\pmod5$, with coefficient $1/16$.
The five equal-value classes are precisely the five supports in Eq.(\ref{vectors}). At the specified parameters, every entry in each support block is $1/16$, proving \eqref{five-product} and separability.
\end{proof}

\begin{figure}[tbp]
\centering\includegraphics[width=0.85\linewidth]{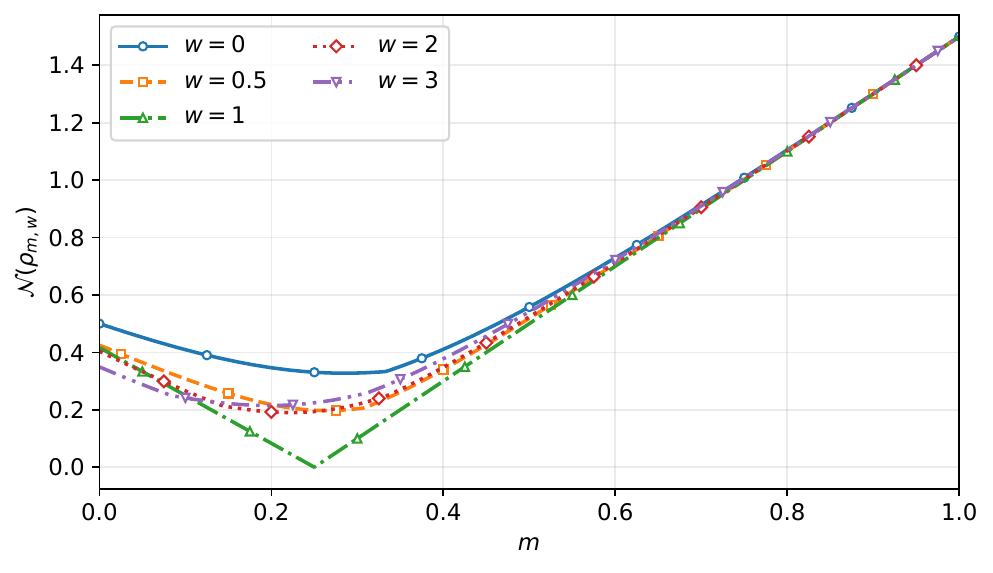}
\caption{Initial negativity for $w=0,0.5,1,2,3$. Only $(m,w)=(1/4,1)$ has zero negativity. Values use the convention
$\mathcal N=(\|\rho^{\PT}\|_1-1)/2$; a maximally entangled four-level pure state has negativity $3/2$.
Curves use distinct line patterns and markers as well as colors.}
\label{Fig:1}
\end{figure}

In Figure(\ref{Fig:1}), we plot negativity vs $m$ for various values of parameter $w$. We observe that negativity is zero only at one point for $m = 0.25$ and non-zero elsewhere. 

\subsection{Three-sector global dephasing}
Define the orthogonal projectors $A=\pr{00}$, $B=\sum_{i=1}^3\pr{ii}$ and $C=\sum_{i\ne j}\pr{ij}$. For $0\leq \gamma \leq1$, set
\begin{align}
J_0&=\gamma (A+B) + C,\notag\\
J_1&=\sqrt{1-\gamma^2}(A -\gamma^2 B),\notag\\
J_2&=(1-\gamma^2)\sqrt{1+\gamma^2}\,B.\label{three}
\end{align}
Thus the last operator has the same nonzero coefficient on
$\ket{11},\ket{22},\ket{33}$. The map
$\mathcal G_\gamma(X)=\sum_{r=0}^2 J_r \, X \, J_r^*$ has sector multipliers
\[
\begin{array}{c|ccc}&A&B&C\\\hline
A&1&\gamma^4&\gamma\\B&\gamma^4&1&\gamma\\C&\gamma&\gamma&1\end{array}.
\]
To verify normalization and this table, take inner products of the coefficient vectors
$(\gamma,\sqrt{1-\gamma^2},0)$,
$(\gamma,-\gamma^2\sqrt{1-\gamma^2},(1-\gamma^2)\sqrt{1+\gamma^2})$ and $(1,0,0)$.
Each has norm one. For $0<\gamma<1$ they are independent, so the minimal Kraus number is three. For $0\leq \gamma<1$, the fixed algebra is $\mathcal B(A\mathcal H)\oplus\mathcal B(B\mathcal H) \oplus\mathcal B(C\mathcal H)$, of dimension $154$. At $\gamma=1$ the channel is the identity and its fixed algebra has dimension $256$.
For $\kappa>0$, $t\geq0$ and $\gamma=e^{-\kappa t}$ the dephasing observable is $G=A-B$, of multiplicities $(1,3,12)$. The Markov generator is $\mathcal L_G(X)=-\kappa[G,[G,X]]$. It is a global diagonal observable, not a sum of local ones: any array $h_{ij}=d_i+e_j$ satisfies $h_{11}+h_{22}=h_{12}+h_{21}$, whereas $G$ gives $-2\ne0$.
\begin{figure}[tbp]
\centering\includegraphics[width=0.85\linewidth]{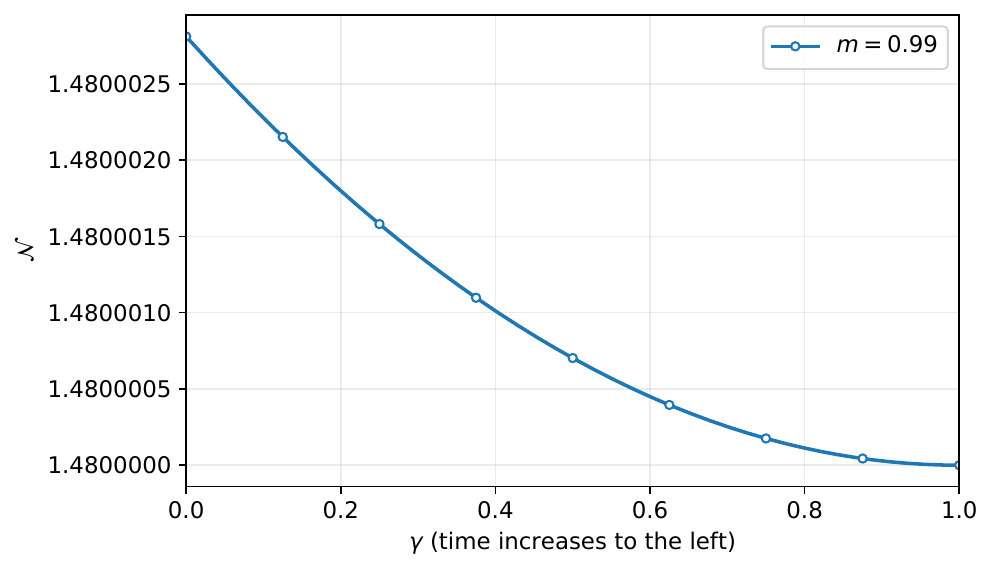}
\centering
\caption{Three-sector global channel at $m=0.99$, $w=1$. The vertical scale resolves the small variation. The constancy of the $\Phi$ component does not by itself make the full state's negativity constant. }
\label{Fig:2}
\end{figure}
The maximally entangled state $\ket\Phi$ lies in the decoherence-free subspace of $\mathcal G_\gamma$. Figure~\ref{Fig:2} tests whether a large weight of this stationary component makes the full state's negativity time-invariant. At $m=0.99$ and $w=1$, the negativity changes slightly but is not constant. It increases as $\gamma$ decreases from one at $t=0$ towards zero in the infinite-time limit.

\begin{figure}[tbp]
\centering\includegraphics[width=0.85\linewidth]{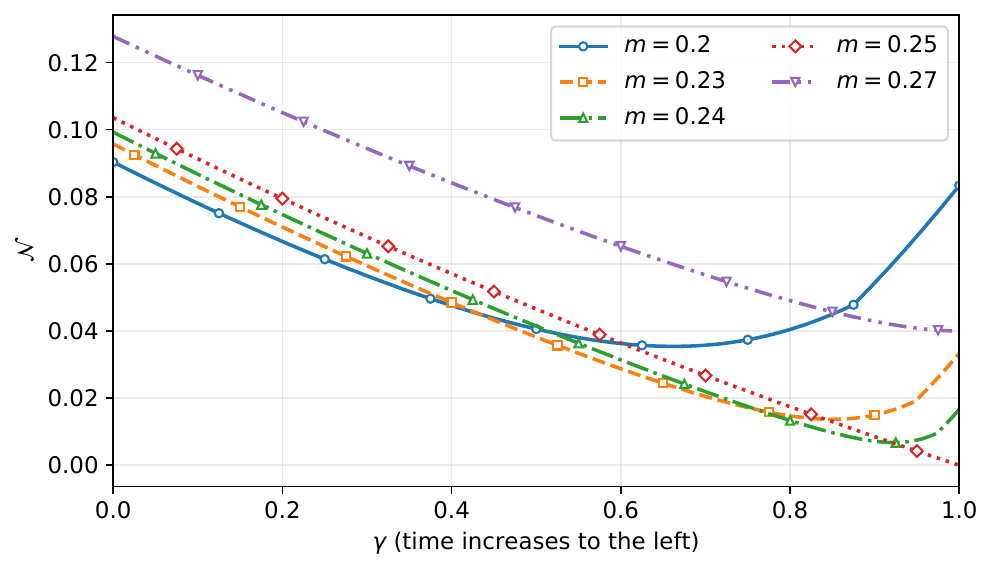}
\caption{Three-sector global channel near $m=1/4$, with $w=1$. The $m=0.25$ curve starts at zero and follows \eqref{creation} exactly.
Time increases from right to left. This global channel can create entanglement.
Curves use distinct line patterns and markers as well as colors.}
\label{Fig:3}
\end{figure}

Figure~\ref{Fig:3} shows the three-sector global channel near $m=1/4$, with $w=1$. The $m=0.24$ trajectory remains NPT, and even the initially separable state at $m=1/4$ becomes entangled. The stationary $\Phi$ component and the nonstationary mixture explain the form of the trajectory, but entanglement is not additive over the mixture. Crucially, $\mathcal G_\gamma$ is generated by the nonlocal observable $G$, rather than by shared local phase rotations. It need not preserve separability and is not subject to the negativity monotonicity proved below for the correlated local channel. The following exact calculation demonstrates this distinction.

At $w=1$, four of the global-channel blocks have the characteristic polynomial of
\begin{eqnarray*}
H_\gamma=\left(\begin{smallmatrix}b&b&a\\b&a&\gamma b\\a&\gamma b&b\end{smallmatrix}\right).
\end{eqnarray*}
The fourth block in \eqref{groups} has eigenvalues $b-a,b-a,a+b-2\gamma b,a+b+2\gamma b$. At the separable point $a=b=1/16$,
\[
\det(zI-H_\gamma)=(z-b(1-\gamma))
\{z^2-b(2+\gamma)z+b^2(\gamma-1)\}.
\]
The quadratic has one negative root for $\gamma<1$. Summing its contribution over the four blocks gives
\begin{equation}\label{creation}
\mathcal N(\mathcal G_\gamma(\rho_{1/4,1}))
=\frac{\sqrt{\gamma^2+8}-2-\gamma}{8}>0\qquad(0\leq \gamma<1).
\end{equation}
In the infinite-time limit $\gamma=0$ this is $(\sqrt2-1)/4$ (non-zero value). 

\subsection{Shared Gaussian collective dephasing}
Let
\begin{equation}\label{local}
D=\diag(0,1,2,3),\quad L=D\otimes I+I\otimes D,\quad
P_s=\sum_{i+j=s}\pr{ij},\quad0\leq s\leq6.
\end{equation}
The sums of labels here are ordinary integers. Equally spaced levels mean that the phase acquired by level $j$ is proportional to $j$. For a spin-$3/2$ system, ordering the magnetic sublevels from $-3/2$ to $3/2$ gives $D=J_z+\tfrac32 I$ in units with $\hbar=1$. The identity shift drops out of the double commutator, so the same dephasing generator describes two such spins exposed to a common fluctuating longitudinal field. Sharing the noise means that both subsystems experience the same phase realization; independent fields would require separate phase averages. For $q=e^{-\kappa t}$,
\begin{equation}\label{shared}
\mathcal E_q(\ket{ij}\bra{k\ell})
=q^{(i+j-k-\ell)^2}\ket{ij}\bra{k\ell}.
\end{equation}
This is the average of conjugations by $e^{-i\theta D}\otimes e^{-i\theta D}$, with $\theta\sim N(0,2\kappa t)$: the local phases are shared, not independent. It preserves separability and cannot increase negativity with time. The phase is a Brownian phase with variance $2\kappa t$, not a static Gaussian frequency multiplied by time; the Markov generator is $\mathcal L_L(X)=-\kappa[L,[L,X]]$.
For the monotonicity assertion, partial transpose converts each local unitary conjugation into another unitary conjugation; convexity of the trace norm then applies. Also $\mathcal E_{q_1}\mathcal E_{q_2}=\mathcal E_{q_1q_2}$.

The seven sector multiplicities are $(1,2,3,4,3,2,1)$. For $0\leq q<1$, the fixed algebra has dimension $44$; at $q=1$ it has dimension $256$. For $0<q<1$ the matrix $(q^{(s-r)^2})_{s,r=0}^6$ is positive definite: its quadratic form is a Gaussian average of a squared trigonometric polynomial, which can vanish only for the zero polynomial. Thus the Choi rank, and minimal Kraus number,
is seven \cite[Theorem~2.22]{watrous}. The channels differ, for example, on $\ket{11}\bra{22}$, fixed by
$\mathcal G_\gamma$ but damped by $q^4$ under $\mathcal E_q$. Both preserve the four-dimensional subspace
$\operatorname{span}\{\ket{03},\ket{12},\ket{21},\ket{30}\}$. On the embedded subspace spanned by $00,01,10,11$, $G=I-L$,
so their restrictions agree when $\gamma=q$. Notice that $\ket{\Phi}$ itself is stationary under $\mathcal G_\gamma$ (it lives in DFS), but not under $\mathcal E_q$: its charges are $1,3,5,3$.

Put $x=q^2$, $a_0=1$, $a_r=\sqrt{\prod_{j=1}^r(1-x^j)}$, $A_s=\sum_{j=0}^{s-1}x^j$, and
\begin{align*}
B_4&=1+x+2x^2+x^3+x^4,\\
B_5&=1+x+2x^2+2x^3+2x^4+x^5+x^6,\\
B_6&=1+x+2x^2+2x^3+3x^4+2x^5+2x^6+x^7+x^8,\\
C_6&=1+x+2x^2+3x^3+3x^4+3x^5+3x^6+2x^7+x^8+x^9.
\end{align*}
An operator-sum representation of \eqref{shared} is
\begin{align}
K_0&=P_0+qP_1+q^4P_2+q^9P_3+q^{16}P_4+q^{25}P_5+q^{36}P_6,\notag\\
K_1&=a_1\,(P_1+qA_2P_2+q^4A_3P_3+q^9A_4P_4+q^{16}A_5P_5+q^{25}A_6P_6),\notag\\
K_2&=a_2\, (P_2+qA_3P_3+q^4B_4P_4+q^9B_5P_5+q^{16}B_6P_6),\notag\\
K_3&=a_3\,(P_3+qA_4P_4+q^4B_5P_5+q^9C_6P_6),\notag\\
K_4&=a_4\,(P_4+qA_5P_5+q^4B_6P_6),\notag\\
K_5&=a_5\,(P_5+qA_6P_6),\notag \\
K_6&=a_6\,P_6.\label{kraus}
\end{align}
Thus $\mathcal E_q(X)=\sum_{r=0}^6K_rXK_r^*$. We provide the factorization and normalization in Appendix~\ref{factor-appendix}.
At $q=1$, $K_0=I$ and the other operators vanish; at $q=0$, we have $K_r=P_r$.

\begin{figure}[tbp]
\centering\includegraphics[width=0.85\linewidth]{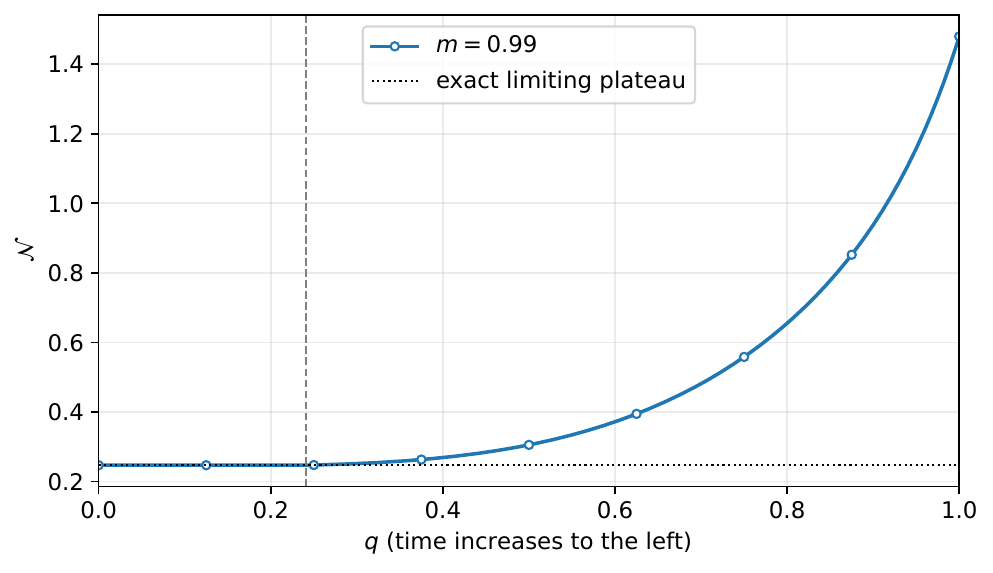}
\centering
\caption{Shared local Gaussian channel, $m=0.99$, $w=1$. For $q\leq q_{\mathrm{fr}}\simeq0.24087974$ the negativity is exactly
$(4m-1)/12$. Time increases from right to left, since $q=e^{-\kappa t}$.}
\label{Fig:4}
\end{figure}
Figure~\ref{Fig:4} shows shared local Gaussian dephasing at $m=0.99$, $w=1$. As $q$ decreases, the negativity reaches a strictly positive constant at a finite time. Theorem~\ref{freezing} gives the exact plateau and its onset.

\begin{figure}[tbp]
\centering\includegraphics[width=0.85\linewidth]{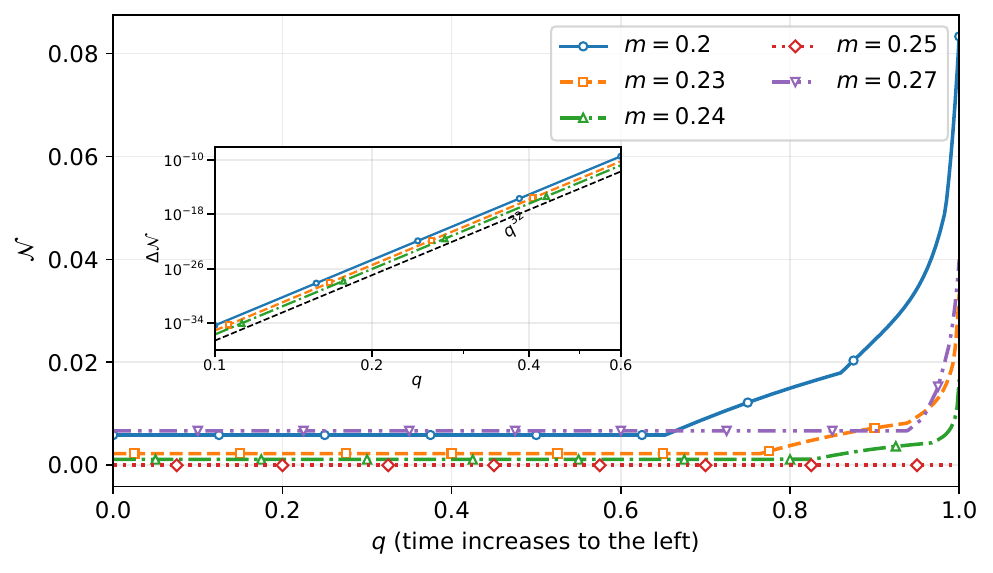}
\caption{Shared local Gaussian channel near $m=1/4$, with $w=1$. The $m=0.27$ curve reaches an exact positive plateau. The curves at
$m=0.20,0.23,0.24$ approach their limiting negativities strictly asymptotically; values indistinguishable at plot resolution need not be equal.
The separable $m=0.25$ trajectory has zero negativity.
The inset plots the positive excess $\Delta\mathcal N=\mathcal N(q)-\mathcal N(0)$ for the three lower weights on logarithmic axes, evaluated with 80 decimal digits; its $q^{32}$ reference slope resolves the asymptotic tail.
Curves use distinct line patterns and markers as well as colors.}
\label{Fig:5}
\end{figure}

Figure~\ref{Fig:5} compares weights near $m=1/4$ under the same channel. The $m=0.27$ trajectory reaches an exact positive plateau. The positive lower-weight curves continue decreasing at every finite time, although their late-time variation is too small to resolve in an ordinary plot. The $m=0.25$ trajectory remains separable.

\section{Exact freezing and asymptotic flattening}\label{dynamics}
Let us define $w=1$, $a=m/4$ and $b=(1-m)/12$. The partial transpose under either diagonal channel has invariant blocks
on the five ordered index sets
\begin{equation}\label{groups}
(00,12,31),\ (01,20,33),\ (02,11,23),\
(03,10,21,32),\ (13,22,30).
\end{equation}
Under $\mathcal E_q$, these give $M_1,M_2,M_3$, up to internal basis permutations, where
\begin{align}
M_1&=\begin{pmatrix}b&qb&q^4a\\qb&a&q^9b\\q^4a&q^9b&b\end{pmatrix},\notag\\
M_2&=\begin{pmatrix}b&q^4a&qb\\q^4a&b&qb\\qb&qb&a\end{pmatrix},\qquad
M_3=\begin{pmatrix}b&q^4a&q^{25}b\\q^4a&b&q^9b\\q^{25}b&q^9b&a\end{pmatrix}.\label{residual}
\end{align}
Specifically, $M_1$ occurs on the ordered bases $(00,12,31)$ and $(33,01,20)$; $M_2$ occurs on $(02,11,23)$ and $M_3$ on $(13,22,30)$.
The four-dimensional block has eigenvector $(0,1,0,-1)$ with eigenvalue $b-a$. On the orthogonal complement, in the basis
$e_1,(e_2+e_4)/\sqrt2,e_3$, it is
\begin{equation}\label{reduced}
C_q=\begin{pmatrix}
b&\sqrt2q^{16}b&q^{16}a\\
\sqrt2q^{16}b&a+b&\sqrt2b\\q^{16}a&\sqrt2b&b
\end{pmatrix}.
\end{equation}
All these formulas follow by substituting \eqref{vectors} into \eqref{shared} and transposing the second pair of indices.
In particular the full decomposition is
\[
\bigl(\mathcal E_q(\rho_{m,1})\bigr)^{\PT}
\cong[b-a]\oplus M_1^{\oplus2}\oplus M_2\oplus M_3\oplus C_q.
\]
There are five complementary three-dimensional blocks and four distinct block types, with $M_1$ occurring twice.
For $a>b$, freezing begins when all complementary blocks become positive semidefinite, leaving only the invariant negative eigenvalue. The density operator continues evolving at every finite time after freezing: its $\ket{01}\bra{12}$ coefficient is $mq^4/4$ and remains time-dependent.

\begin{lemma}\label{strictness}
Let $\rho_0$ be any density operator on $\mathbb C^4\otimes\mathbb C^4$ and $\rho(t)=\mathcal E_{e^{-\kappa t}}(\rho_0)$, with $\kappa>0$. If its negativity has the same value at two distinct finite times, that value equals its infinite-time limit and remains constant thereafter.
\end{lemma}
\begin{proof}
After partial transpose the Gaussian conjugations have generator $D\otimes I-I\otimes D$. Put $H=\rho(t)^{\PT}$ and choose a Hermitian
contraction $Z$ attaining the trace-norm dual formula for $\rho(t+\tau)^{\PT}$, $\tau>0$.
Each conjugation $V_\theta H V_\theta^*$ satisfies $\Tr(ZV_\theta HV_\theta^*)\leq\|H\|_1$.
Equality of the trace norms at the two times forces equality in the averaged dual bound, and hence equality for every $\theta$, because the Gaussian density is everywhere positive and the integrand
is continuous. Average this equality uniformly over $[0,2\pi]$. The generator has integer eigenvalues, so this average is exactly the
limiting charge pinching $H_\infty$. It follows that
$\|H\|_1=\Tr(ZH_\infty)\leq\|H_\infty\|_1\leq\|H\|_1$.
\end{proof}

\begin{theorem}\label{freezing}
For the shared local channel \eqref{shared}, the following hold.
\begin{enumerate}
\item If $1/4<m<1$, there is $q_{\mathrm{fr}}(m)\in(0,1)$ such that
\[
\mathcal N(\mathcal E_q(\rho_{m,1}))=a-b=\frac{4m-1}{12}
\quad\Longleftrightarrow\quad0\leq q\leq q_{\mathrm{fr}}(m).
\]
Exactly, $q_{\mathrm{fr}}$ is the largest $q\in[0,1]$ for which $M_1,M_2,M_3,C_q$ are all positive semidefinite. Negativity decreases strictly with time before this plateau.
\item If $0\leq m<1/4$, then, for every $0<q\leq1$,
\[
\mathcal N(\mathcal E_q(\rho_{m,1}))>
\mathcal N(\mathcal E_0(\rho_{m,1}))
=\frac{\sqrt{a^2+8b^2}-a-2b}{2}>0.
\]
It decreases strictly with time at every finite time.
\item At $m=1/4$ the negativity is identically zero.
At $m=1$ it equals $1/4+q^4+q^{16}/4$.
\end{enumerate}
Thus a positive plateau equal to the infinite-time limit is reached at finite time exactly for $1/4<m<1$.
\end{theorem}
\begin{proof}
Suppose $a>b>0$. The eigenvalue $b-a<0$ contributes $a-b$ at every $q$. At $q=0$ the three matrices in \eqref{residual} are positive diagonal matrices. The remaining $2\times2$ part of $C_0$ is
\[
T=\begin{pmatrix}a+b&\sqrt2b\\\sqrt2b&b\end{pmatrix},
\quad \det T=b(a-b)>0,
\]
so $C_0$ too is positive definite. The least eigenvalues depend continuously on $q$, hence all four complementary matrices remain
positive on some interval $[0,\varepsilon]$. There the only negative eigenvalue is $b-a$. At $q=1$, $M_2$ has eigenvalue $b-a$ on $(1,-1,0)$, so negativity is strictly greater than $a-b$ there. Moreover $\mathcal N(\mathcal E_q(\rho))$ is nondecreasing in $q$ by the local
random-unitary semigroup property. Its level set at its lower bound $a-b$ is consequently an initial interval. Continuity makes this interval
closed. Its upper endpoint belongs to $(0,1)$ and is characterized by positivity of all four complementary blocks, as claimed.
The freezing time is $t_{\mathrm{fr}}=-\log q_{\mathrm{fr}}/\kappa<\infty$. Lemma~\ref{strictness} gives strict decay before its onset.

Now $0\leq a<b$. At $q=0$ all other blocks are positive semidefinite, whereas $T$ has one negative eigenvalue
$\lambda_-=(a+2b-\sqrt{a^2+8b^2})/2$. Write $v=(\sqrt2b,a)^T$. If $u_-$ is a unit negative eigenvector of $T$,
then $v^Tu_-\ne0$: otherwise $v$ would be a positive eigenvector, but
\[
\det[\,v,Tv\,]=2\sqrt2b(b^2-a^2)\ne0.
\]
The compression of $C_q$ to the span of $e_1$ and $(0,u_-)$ is
$\left(\begin{smallmatrix}b&q^{16}v^Tu_-\\q^{16}v^Tu_-&\lambda_-\end{smallmatrix}\right)$.
For $q>0$ its smaller eigenvalue is strictly less than $\lambda_-$. The variational principle gives
$\lambda_{\min}(C_q)<\lambda_-$; its contribution alone proves the strict negativity inequality. This includes $a=0$.
Lemma~\ref{strictness} again supplies strict decay at finite times.

At $m=1/4$, Proposition~\ref{ppt-classification} and preservation of separability prove the assertion. At $m=1$, dephasing the pure initial state gives a generally mixed maximally correlated density operator with six unordered coherence pairs: four of weight $q^4/4$, one of weight $q^{16}/4$, and one of weight $1/4$. Each gives the corresponding negative partial-transpose eigenvalue, which proves the last formula.
\end{proof}

The exact boundary criterion requires only positive semidefiniteness of four specified $3\times3$ matrices, equivalently their principal minors.
Numerical bisection gives $q_{\mathrm{fr}}(0.27)\simeq0.93864509$ and $q_{\mathrm{fr}}(0.99)\simeq0.24087974$. Exact rational
principal-minor comparisons give the enclosures
\[
0.93864508<q_{\mathrm{fr}}(0.27)<0.93864510,\qquad
0.24087973<q_{\mathrm{fr}}(0.99)<0.24087976.
\]
In both enclosures every complementary principal minor is positive at the lower endpoint, while $\det M_1<0$ at the upper endpoint.
The determinant used at the upper endpoints is explicitly
\[
\det M_1=ab^2(1+2q^{14})-a^3q^8-b^3(q^2+q^{18}).
\]
For each fixed $0<m<1/4$, the other blocks are positive definite near $q=0$. Put $\varepsilon=q^{16}$ and write
\[
C(\varepsilon)=\begin{pmatrix}b&\varepsilon v^T\\\varepsilon v&T\end{pmatrix}.
\]
At $\varepsilon=0$, its negative eigenvalue $\lambda_-$ is simple, with unit eigenvector $(0,u_-)$. The first-order perturbation has zero expectation in this vector and couples it only to the eigenvector $e_1$ of eigenvalue $b$. Thus the second-order coefficient is $|v^Tu_-|^2/(\lambda_--b)$. Conjugation by $\diag(-1,1,1)$ sends $C(\varepsilon)$ to $C(-\varepsilon)$, so the analytic eigenvalue branch is even:
\[
\lambda_-(\varepsilon)=\lambda_-
-\frac{|v^Tu_-|^2}{b-\lambda_-}\varepsilon^2+O(\varepsilon^4).
\]
The coupling is nonzero by the determinant calculation in the proof of Theorem~\ref{freezing}. Since the remaining eigenvalues are positive for small $q$, changing the sign of this eigenvalue gives
\[
\mathcal N(\mathcal E_q(\rho_{m,1}))-
\mathcal N(\mathcal E_0(\rho_{m,1}))
=c(m)q^{32}+O(q^{64}),\qquad
c(m)=\frac{|v^Tu_-|^2}{b-\lambda_-}>0.
\]
At $m=0$, additional zero eigenvalues of the residual blocks instead give $\mathcal N(q)-\mathcal N(0)=q^2/3+O(q^4)$: the two $M_1$ blocks contribute $2bq^2+O(q^4)$ and $M_2$ contributes $2bq^2+O(q^4)$, with $b=1/12$; the remaining contributions have higher
order. This endpoint is still strictly asymptotic. The $q^{32}$ tail explains the flat appearance of the positive low-weight curves without
replacing asymptotic convergence by exact equality.

\section{Comparison with known circulant and PPT-entangled states}
The comparisons in this section distinguish the initial family from known PPT-entangled constructions and identify a connection through a noninvertible local phase average. This relation is a local-channel image, rather than equivalence under invertible local transformations.

In this section, we compare our states with some of the known entangled states in $4 \otimes 4$ quantum system. By defining all basis labels as modulo four, let us write $S\ket j=\ket{j+1}$, $\ket\Omega=\frac12\sum_i\ket{ii}$, $P=\pr\Omega$ and 
$\sigma_k=\frac14\sum_i\pr{i,i+k}$. The family of quantum states defined in \cite[Eq.~(3.43)]{jba} can be written as 
\begin{equation}\label{jfamily}
\rho_{\beta,\alpha} =\frac{3P+\beta\sigma_1+\alpha\sigma_2+(10-\beta)\sigma_3}{13+\alpha},
\quad0\leq\beta\leq10,\quad\alpha\geq0.
\end{equation}
In fact, these states are just a slice of the circulant class treated in \cite[Theorem~1]{ck}. Let $\zeta=e^{2\pi i/3}$ and define
\begin{align*}
U_\alpha&=\diag(1,\zeta^{\alpha_1},\zeta^{\alpha_2},\zeta^{\alpha_3}),\\
\mathcal T(X)&=27^{-1}\sum_{\alpha\in\{0,1,2\}^3}
(U_\alpha\otimes\overline U_\alpha)X(U_\alpha\otimes\overline U_\alpha)^*.
\end{align*}
Then
\begin{equation}\label{twirl}
\mathcal T((I\otimes S^{-1})\rho_{m,w}(I\otimes S))
=mP+\frac{1-m}{w^2+2}(\sigma_1+\sigma_2+w^2\sigma_3).
\end{equation}
To see this, the phase average retains a matrix unit $\ket{ij}\bra{k\ell}$ exactly when the multisets $\{i,\ell\}$ and $\{j,k\}$ coincide.
Each independent exponent is between $-2$ and $2$, so the cube-root average vanishes unless it is zero. The shift sends maximally entangled state $\ket{\Phi}$ to maximally entangled state $\ket{\Omega}$; each shifted $\chi_j$ has one amplitude $w$ in shift three and unit amplitudes in shifts one and two, with the common normalization in \eqref{vectors}. Their off-diagonal terms disappear, proving \eqref{twirl}. Coefficient matching shows that its image belongs to \eqref{jfamily} exactly when
\begin{equation}\label{intersection}
\beta=\alpha=\frac{10}{1+w^2},\qquad
m=\frac{3(1+w^2)}{23+13w^2}.
\end{equation}
At $(m,w)=(1/4,1)$, the image is $(P+\sigma_1+\sigma_2+\sigma_3)/4$, a separable rank-thirteen state.
Explicitly, it is the average of
$\pr{u_\alpha}\otimes\pr{\overline u_\alpha}$ with
$\ket{u_\alpha}=\frac12\sum_{i=0}^3\zeta^{\alpha_i}\ket i$ and $\alpha_0=0$.
The same matrix-unit calculation proves this product decomposition. This balanced circulant state's separability is also the established case
of \cite[Theorem~1]{ck}; the rank-five certificate \eqref{five-product} concerns the original state before averaging.
The twirl annihilates, for example, $\ket{01}\bra{02}$ and is therefore noninvertible on the full operator space.

PPT in \eqref{jfamily} requires $\beta(10-\beta)\geq9$ and $\alpha \geq3$, from its $2\times2$ blocks. Thus $1\leq\beta\leq9$ and all three $\sigma_k$ weights are positive. Every PPT member has rank thirteen, excluding local-unitary or invertible
local-filter equivalence to any member of \eqref{state}, whose rank is at most five. This rank statement concerns invertible filters, not arbitrary local quantum operations such as \eqref{twirl}.

Let us compare our states with another family of quantum states \cite{tv} defined as 
\begin{equation*}
\tau_{p,r}=p\sum_{j=1}^4\pr{\psi_j}+r\sum_{j=5}^6\pr{\psi_j},
\end{equation*}
where $p,r\geq0$, $4p+2r=1$, and with
\begin{align*}
\psi_1&=(\ket{01}+\ket{23})/\sqrt2,&
\psi_2&=(\ket{10}+\ket{32})/\sqrt2,\\
\psi_3&=(\ket{11}+\ket{22})/\sqrt2,&
\psi_4&=(\ket{00}-\ket{33})/\sqrt2,\\
\psi_5&=(\ket{03}+\ket{12})/2+\ket{21}/\sqrt2,&
\psi_6&=(-\ket{03}+\ket{12})/2+\ket{30}/\sqrt2.
\end{align*}
Its partial-transpose spectrum is
\[
\left\{\left(\frac p2\pm\frac r{2\sqrt2}\right)^{\!\times4},
\left(\frac r2\pm\frac p{\sqrt2}\right)^{\!\times2},0^{\times4}\right\}.
\]
Consequently its unique PPT point is $p_0=(2-\sqrt2)/4$, $r_0=(\sqrt2-1)/2$. This is the rank-six PPT-entangled state of T\'oth and V\'ertesi
\cite[corrected Supplemental Material, p.~3]{tv}. In the displayed basis it equals its partial transpose and its realignment
trace norm is $5/2-\sqrt2>1$. It therefore differs in both rank and separability from \eqref{five-product}.

\section{Exact realignment-moment detection}
We evaluate the moment criterion exactly on the initial family and compare it with ordinary realignment. On $w=1$, this comparison identifies precisely the entangled states missed by the moment bound despite detection by ordinary realignment.
Define $R(\rho)_{ik,j\ell}=\rho_{ij,k\ell}$ and $T_1=\Tr(R^*R)$. The realignment rank $k=\rank R$ is distinct from
the density-matrix rank. Let $D_k$ be the product of the $k$ positive eigenvalues of $R^*R$. The criterion in
\cite[Theorem~1, Eq.~(23)]{aam} detects entanglement when
\begin{equation}\label{criterion}
\mathcal R=k(k-1)D_k^{1/k}+T_1-1>0.
\end{equation}
Indeed, for the positive singular values $s_i$ the arithmetic--geometric mean inequality gives
$\|R\|_1^2=T_1+2\sum_{i<j}s_is_j\geq T_1+k(k-1)(\prod_i s_i^2)^{1/k}$. For a separable state $\|R\|_1\leq1$ \cite{rudolph,chen}.
This is a sufficient entanglement criterion; failure to detect does not imply separability.

For \eqref{state}, return to $a=m/4$, $b=(1-m)/(4(w^2+2))$, and introduce the following scalar quantities for this section only. In particular, $C_0$ below is a scalar, not the matrix $C_q$ at $q=0$:
\begin{align*}
T&=m^2+(1-m)^2/4,& A_0&=(a-bw^2)^2,\\
C_0&=(a-b)^2+b^2(w^2-1)^2,& E_0&=(a-bw)^2,\\
H_0&=a^2+abw-2b^2w,& L_0&=2a^2+2abw+3b^2w^2+2b^2.
\end{align*}
For $m<1$ the complete rank table is
\begin{center}
\begin{tabular}{@{}lll@{}}\toprule
Condition & $k$ & $D_k$\\\midrule
$A_0E_0H_0\ne0$ &16&$A_0C_0^2E_0^4H_0^8/16$\\
$A_0=0$ &15&$C_0^2E_0^4H_0^8/16$\\
$E_0=0$ &12&$A_0C_0^2H_0^8/16$\\
$H_0=0$ &12&$A_0C_0^2E_0^4L_0^4/16$\\
$(m,w)=(0,0)$ &7&$2^{-34}$\\
$(m,w)=(1/4,1)$ &5&$(1/16)(9/256)^4$\\\bottomrule
\end{tabular}
\end{center}
The first four rows exclude the last two points. The exact detected region is
\begin{equation}\label{region}
[k(k-1)]^kD_k>(1-T)^k.
\end{equation}
Equality is undetected. All states at $m=1$ are detected with $\mathcal R=15$.

\begin{proof}
Permuting realignment rows and columns gives one circulant block
\[
M=\begin{pmatrix}bw^2&a&b&b\\b&bw^2&a&b\\b&b&bw^2&a\\a&b&b&bw^2\end{pmatrix}
\]
and four $3\times3$ blocks with the squared-singular-value polynomial of
$N=\left(\begin{smallmatrix}bw&a&b\\bw&bw&a\\a&bw&b\end{smallmatrix}\right)$.
The row/column index sets for these four blocks are
\[
\begin{gathered}
(01,13,32;03,12,20),\quad (02,21,30;01,13,32),\\
(03,12,20;10,23,31),\quad (10,23,31;02,21,30).
\end{gathered}
\]
The block $M$ uses rows and columns $00,11,22,33$.
Since $a+b(w^2+2)=1/4$, the squared singular values of $M$ are
$1/16,A_0,C_0,C_0$. Direct multiplication gives
\[
\det(zI-N^*N)=(z-E_0)(z^2-L_0z+H_0^2).
\]
For $E_0H_0\ne0$, one such block has positive-eigenvalue product
$E_0H_0^2$; when only $E_0$ vanishes, the product is $H_0^2$;
when only $H_0$ vanishes, it is $E_0L_0$.
When both vanish the rank is one with eigenvalue $L_0$.
For $b>0$, $E_0=H_0=0$ implies $a=bw$ and
$2b^2w(w-1)=0$, giving precisely the two exceptional points.
Likewise $C_0=0$ occurs exactly at $a=b,w=1$; intersections involving
$A_0=0$ give only those same points, because substitution $a=bw^2$
gives $H_0=b^2w(w-1)(w^2+2w+2)$.
This proves exhaustiveness and disjointness of the table.
At $(0,0)$ the positive squared singular values are $1/16$ and
$1/32$ six times; at $(1/4,1)$ they are $1/16$ and $9/256$ four times.
Multiplication yields the displayed $D_k$.
Orthonormality of the defining vectors gives $T_1=\Tr\rho^2=T$.
For $m<1$, $1-T>0$, so taking the $k$th power in \eqref{criterion}
is equivalent to \eqref{region}. At $m=1$ all sixteen singular values
are $1/4$, giving $\mathcal R=15$.
\end{proof}

On $w=1$, outside the separable point, the singular values are
$1/4$, $|4m-1|/12$ eleven times and $(m+2)/12$ four times.
Consequently detection is equivalent to
\begin{equation}\label{slice}
2^{32}5^{16}(4m-1)^{22}(m+2)^8
>3^{14}(1-m)^{16}(5m+3)^{16}.
\end{equation}
For the ratio $F$ of the two sides,
\[
(\log F)'=\frac{8(5m^3+49m^2+19m+71)}
{(1-m)(m+2)(4m-1)(5m+3)}.
\]
Thus $F$ strictly decreases below $1/4$ and increases above it;
$F(0)>1$, $F(1/4)=0$, and $F\to\infty$ at $1$.
There are exactly two equality roots, approximately $0.14985684$ and
$0.34497271$. Detection occurs below the first or above the second.
For example, $m=1/5$ and $m=3/10$ are NPT but undetected.
By comparison, summing the same singular values gives the ordinary
realignment norm, including the rank-drop point,
\[
\|R(\rho_{m,1})\|_1
=\frac14+\frac{11|4m-1|}{12}+\frac{m+2}{3}
=\begin{cases}(11-20m)/6,&0\leq m\leq1/4,\\
4m,&1/4\leq m\leq1.
\end{cases}
\]
It exceeds one exactly when $m\ne1/4$, so ordinary realignment detects
every entangled initial state on this slice. The missed intervals thus
measure the loss in the moment bound \eqref{criterion}, not a failure
of ordinary realignment. The rank-sixteen generic calculation is an
exact evaluation, without an assertion of low-order experimental efficiency.
For the comparison state $\tau_{p_0,r_0}$ the eight positive singular
values are $1/4$, $(3-2\sqrt2)/4$, $(\sqrt2-1)/4$ twice and
$(2-\sqrt2)/4$ four times. Their sum gives the trace norm above and
\eqref{criterion} gives $(17\sqrt2-24)/2>0$, reproducing
\cite[Section~IV.B.1]{aam}.

\section{Conclusions}
Shared local Gaussian dephasing gives a positive finite-time negativity plateau on the equal-amplitude slice exactly for $1/4<m<1$, with value $(4m-1)/12$. For $0\leq m<1/4$, negativity is strictly decreasing at every finite time and remains above its positive limiting value. The separable point $m=1/4$ and the pure-state endpoint $m=1$ complete the classification. Thus the same family and noise generator exhibit both exact post-transient freezing and strictly asymptotic convergence.

The spectral proof distinguishes these regimes by the signs of five complementary three-dimensional blocks and an invariant partial-transpose eigenvalue. The equality-at-two-times lemma excludes a temporary flat interval before further decay; positivity of the complementary blocks gives the exact onset criterion. The perturbative excess is of order $q^{32}$ for fixed $0<m<1/4$, compared with $q^2$ at $m=0$. This difference explains why ordinary plots can conceal strict convergence.

The initial-state separability certificate, explicit Gaussian Kraus operators, known-state comparisons and evaluation of the realignment-moment criterion support the dynamical result. The global-versus-correlated-local distinction and entanglement creation by the global map are inherited from the earlier qutrit analysis \cite{zam}; here their realization on the four-level family provides a comparison with the additive local dynamics. Ordinary realignment detects every entangled initial state at $w=1$, so the states missed by the moment criterion quantify the loss in that existing bound. Extending the freezing-boundary analysis away from $w=1$ remains a natural question; the initial-state classification already covers $w\geq0$.

\paragraph{Mathematical coverage.}
The continuous parameter statements are established by the written proofs.
The accompanying programs check the product-state identity, channel
polynomials, partial-transpose and realignment block identities, rank
examples and boundary sign comparisons by exact arithmetic. The main plots and quoted boundary approximations use double-precision numerical eigenvalues; the logarithmic excess-negativity inset uses 80-decimal-digit arithmetic.

\section*{Statements and Declarations}
\paragraph{Funding.}
The authors received no specific funding for this work.

\paragraph{Competing interests.}
The authors declare no competing interests.

\paragraph{Data and code availability.}
The reproducibility archive \texttt{source.zip} is submitted as supplementary material with this manuscript for referee access. It contains the LaTeX and BibTeX sources, vector figures, figure data, verification and plotting scripts, pinned requirements and a README. The calculations use Python 3.13.13, NumPy 2.5.5, SymPy 1.14.0, Matplotlib 3.11.2 and mpmath 1.3.0. Run \texttt{python3 verify.py} for the exact identities, rational boundary checks and 1,010 numerical comparisons; \texttt{python3 equal\_amplitude\_checks.py} for the equal-amplitude identities; and \texttt{python3 plot\_figures.py} to regenerate all five figures and their data. The boundary checks use exact rational endpoints $0.93864508,0.93864510$ at $m=27/100$ and $0.24087973,0.24087976$ at $m=99/100$: all complementary principal minors are positive at each lower endpoint and $\det M_1<0$ at each upper endpoint. The README gives the environment settings and the clean \texttt{pdflatex}/\texttt{bibtex} build sequence.

\paragraph{AI declaration.}
In preparing this manuscript, the authors used AI tools (ChatGPT Pro, Codex and Claude) for literature search, computation and checking. All proofs can be checked by hand, and the computations can be reproduced from the accompanying verification package. The authors take full responsibility for the content.

\appendix
\section{Gaussian factorization and endpoint normalization}\label{factor-appendix}
Let $G_{nr}(x)$ be the Gaussian binomial polynomial, defined by
$G_{00}=1$, $G_{nr}=0$ outside $0\leq r\leq n$, and
\[
G_{nr}(x)=G_{n-1,r-1}(x)+x^rG_{n-1,r}(x).
\]
These standard polynomials \cite[Section~17.2(ii), Eqs.~(17.2.27), (17.2.31)]{dlmf} satisfy
$G_{n1}=A_n$, $G_{42}=B_4$, $G_{52}=B_5$, $G_{62}=B_6$ and
$G_{63}=C_6$. For $0\leq s,r<N$ set
\begin{equation}\label{factor}
b_{sr}=\begin{cases}
q^{(s-r)^2}G_{sr}(q^2)\sqrt{\prod_{j=1}^r(1-q^{2j})},&r\leq s,\\
0,&r>s.
\end{cases}
\end{equation}
We claim $\sum_rb_{sr}b_{ur}=q^{(s-u)^2}$.
This is the normalized Rogers--Szeg\H{o} coefficient factorization
\cite[Proposition~4.13]{js}; here is a direct proof.
The Newton expansion
\[
z^u=\sum_{r=0}^uG_{ur}(R)\prod_{j=0}^{r-1}(z-R^j)
\]
follows by induction: multiplication by $z$ sends the $r$th product
to the next product plus $R^r$ times itself, giving the defining recurrence.
For $0<x=q^2<1$ take $R=x^{-1}$, $z=x^{-s}$.
The product formula for $G$, obtained from the same recurrence, gives
$G_{ur}(x^{-1})=x^{-r(u-r)}G_{ur}(x)$ and
\[
\prod_{j=0}^{r-1}(x^{-s}-x^{-j})
=x^{-sr}G_{sr}(x)\prod_{j=1}^r(1-x^j).
\]
Terms with $r>s$ vanish. Multiplication by $x^{su}$ therefore yields
\[
1=\sum_{r=0}^{\min(s,u)}x^{(s-r)(u-r)}G_{sr}(x)G_{ur}(x)
\prod_{j=1}^r(1-x^j).
\]
Multiplying by $q^{(s-u)^2}$ proves the claim, since
$(s-u)^2+2(s-r)(u-r)=(s-r)^2+(u-r)^2$.
Continuity extends it to $q=0,1$, with $q^0=1$ there.
For $N=7$, $K_r=\sum_sb_{sr}P_s$ is exactly \eqref{kraus}; hence
\[
\sum_rK_rXK_r^*=\sum_{s,u}q^{(s-u)^2}P_sXP_u,
\qquad \sum_rK_r^*K_r=\sum_sP_s=I.
\]
For $0<q<1$ the triangular diagonal $b_{rr}>0$ proves rank seven.
At $q=0$ it is the identity factor and at $q=1$ only its first column
is nonzero, proving the endpoint assertions.

\bibliographystyle{unsrt}
\bibliography{references}
\end{document}